\documentclass[onecolumn,showkeys,preprintnumbers,amsmath,amssymb,floatfix]{revtex4}

\usepackage{graphicx}
\usepackage{dcolumn}
\usepackage{bm}

\newcommand{\beq}{\begin{equation}}
\newcommand{\eeq}{\end{equation}}
\newcommand{\bey}{\begin{eqnarray}}
\newcommand{\eey}{\end{eqnarray}}

\begin{document}

\title{Gravitational collapse in spatially isotropic coordinates}

\author{Megan Govender}
\email{govenderm43@ukzn.ac.za} \affiliation{Astrophysics and Cosmology Research Unit, School of Mathematical Sciences, University of KwaZulu Natal, Durban, 4041,
South Africa.}

\author{Robert Bogadi}
\email{213572912@stu.ukzn.ac.za} \affiliation{Astrophysics and Cosmology Research Unit, School of Mathematical Sciences, University of KwaZulu Natal, Durban, 4041,
South Africa.}

\author{Ranjan Sharma}
\email{rsharma@iucaa.ernet.in}\affiliation{Department of Physics,
P. D. Women's College, Jalpaiguri 735101, India.}

\author{Shyam Das}
\email{dasshyam321@gmail.com}\affiliation{Department of Physics,
P. D. Women's College, Jalpaiguri 735101, India.}

\date{\today}

\begin{abstract}
We investigate the dynamical nature of the collapse process of a spherically symmetric star in quasi-static hydrodynamical equilibrium. The star collapses from an initial static configuration by dissipating energy in the form of a radial heat flux. The dissipation ensures that the singularity is never formed and the stellar mass completely evaporates over a finite time.
\end{abstract}

\keywords{Gravitational collapse; Exact solutions; Einstein's field equations; Radiating star model; Extended irreversible thermodynamics.}

\maketitle

\section{\label{sec1}Introduction}

Seeking exact solutions to Einstein's field equations, capable of describing realistic astrophysical systems, has been an area of active research ever since the discovery of the Schwarzschild solution in 1916\cite{Shapiro}. Various techniques and assumptions based on particle physics, hydrostatic equilibrium and physical observations have been employed to generate reasonably viable stellar models. At the same time, various ad-hoc approaches have also been used to simplify the non-linearity of the field equations so as to generate solutions which are well behaved and can be utilized to describe realistic physical processes. A comprehensive study of exact static solutions of the Einstein field equations, based on physically acceptability, picks out a very small class of solutions that satisfy all the conditions for hydrostatic equilibrium and causality\cite{Finch,Delgaty}. Nevertheless, in the absence of any reliable information about the physics of matter content at extremely high density, a geometrical approach has been found to be a meaningful technique to study compact stellar objects, a `natural laboratory' for understanding particle interactions at extreme conditions. For example, the Tikekar super-dense stellar model\cite{Tikekar}, describing the gravitational field of a highly compact spherically symmetric star, was shown to exhibit a reasonable EOS for neutron stars. Later, this observation has  prompted many investigators to look for exact solutions capable of describing a large variety of astrophysical systems where relativistic effects can not be ignored; thereby expanding the Tikekar model to include charge\cite{Sharma01}, pressure anisotropy\cite{Karma}, quark matter\cite{Sharma02}, scalar fields\cite{Sharma03} and higher dimensional analogues\cite{singe1,singe2}. Amongst many such realistic models, the static stellar model proposed by Pant and Sah\cite{Pant} is of particular interest. The Pant and Sah\cite{Pant} model describes a spherically symmetric compact star in spatially isotropic coordinates. The solution regains the well-known Buchdahl polytrope solution\cite{Buchdahl} of index $5$. The physical viability of the Pant and Sah\cite{Pant} model was recently looked at in detail by Deb {\em et al}\cite{Deb} and it has been shown that the model can be utilized to describe a wide variety of compact stellar objects including strange stars.

In this paper, we have incorporated dynamical effects into the Pant and Sah\cite{Pant} model by allowing certain model parameters to evolve with time. This has allowed us to investigate the non-adiabatic collapse of a star in a spatially isotropic back ground space-time. In our time-dependent model, we have assumed that the star begins it collapse from an initial static configuration by dissipating energy in the form of a radial heat flux. One regains the Pant and Sah\cite{Pant} model as the  static limit of dynamical collapse process.

Although the issue of gravitational collapse was first taken up by Oppenheimer and Snyder\cite{OppenS}, study of a more realistic collapsing scenario in the presence of dissipative processes was possible only when Vaidya\cite{Vaidya} provided the metric corresponding to the exterior gravitational field of a radiating star. A formal treatment of the junction conditions required for the smooth matching of the collapsing core to the exterior non-empty space-time was provided by Santos\cite{Santos}. These junction conditions provided the impetus for studying dissipative collapse, with much of the early work done by Herrera and co-workers\cite{Herrera1,Herrera2,Herrera3,Herrera4,Herrera5,Herrera6,Herrera7,Herrera8,Herrera9}. For a radiating collapsing star, the pressure at the boundary is proportional to the magnitude of the heat flux and hence this gives rise to a temporal evolution equation for the metric functions. As in the case of static model, various solutions for radiating stars have been found based on physics, dynamical stability and ad-hoc assumptions. The interior space-time of these models have been generalized to include (apart from heat flow) the anisotropic pressure, bulk viscosity, shear and the electromagnetic field\cite{Bonnor,Oliveira1,Oliveira2,Prisco1,Prisco2,Pinheiro1,Chan1,Chan2,Chan3,Tikekar1,Maharaj1,Sharif1,Sharif2,Sharif3,Sharif4,Sharif5,Barreto1,Ghezzi1,Goswami1,Thiru1,Govinder1,Govender1,Goenner1,Sarwe1,Sharma1,t1g1,mks}. A comprehensive review of various approaches and analyses involving gravitationally collapsing systems may be found in Ref.~\cite{Joshi01}. An interesting approach was adopted by Kramer\cite{Kramer1} in which the interior Schwarzschild solution was written in isotropic coordinates and the mass parameter was allowed to become time-dependent. The radiating Schwarzschild-like solution had as its source term, a perfect fluid with heat flow. Since the interior was radiating energy, the exterior was non-empty and was described by Vaidya's outgoing metric. Kramer provided a first integral of the boundary condition required for the matching of the interior to the Vaidya solution. Maharaj and Govender\cite{Maharaj2} presented the full temporal behaviour of the Kramer model in terms of Li integrals. The complicated form of the analytical solution for the temporal behaviour did not warrant a full study of the physics of the model. The present work takes up the initiative to use the Kramer algorithm to provide a full descriptive model of dissipative gravitational collapse.

As the collapse process begins in a massive star after exhausting all its thermonuclear fuel, prediction of the final stage of the collapsing star becomes very much speculative in nature\cite{Chandra}. In fact, one of the the most outstanding challenges in general relativity has been the prediction of the end state of a gravitational bounded system.  In the context of Cosmic Censorship Conjecture(CCC), the general relativistic prediction is that such a collapse must terminate in a black hole; though there are several counter examples where it has been shown that a naked singularity is more likely to be formed\cite{Joshi01}. In our dynamical model, we show that as the star begins its collapse from an initial static configuration with acceptable physical conditions, the dissipation process ensures that the singularity is never formed. Actually, the stellar mass completely evaporates over a finite time.

Our paper has been organized as follows: In Sec.~\ref{sec2}, we have laid down the equations governing collapse in spherically symmetric and spatially isotropic coordinates. For a radiating star, the exterior space-time is appropriately described by the Vaidya metric\cite{Vaidya} and the junctions conditions joining the interior and the exterior regions have been obtained in this section. By introducing a time dependent variable in the Pant and Sah\cite{Pant} model, we have developed a dynamical model of the radiating collapsing star in Sec.~\ref{sec3}. In Sec.~\ref{sec4}, we have studied the physical behaviour of the evolving star and in Sec.~\ref{sec5} we have investigated the thermodynamics of the collapsing star within the framework of extended irreversible thermodynamics. Some concluding remarks have been made in Sec.~\ref{sec6}.

\section{\label{sec2} Interior and exterior space-times:}
We write the interior space-time of a spherically symmetric shear-free collapsing star in spatially isotropic coordinates as
\begin{equation}
ds_{-}^2 = -A^2(r,t) dt^2 + B^2(r,t)[dr^2 + r^2 (d \theta^2 + \sin^2 \theta d \phi^2)].\label{intm1}
\end{equation}
We assume that the material composition filling the interior of the collapsing object is a perfect fluid with out-flowing radiation and accordingly we express the energy-momentum tensor in the form
\begin{equation}
T_{ij} = (\rho + p)u_i u_j + p g_{ij} + q_i u_j + q_j u_i,\label{emt1}
\end{equation}
where $\rho$ is the energy density, $p$ is the isotropic fluid pressure, $ u^{i} = (1/A)\delta^i_0$ is the time-like $4$-velocity of the
fluid and $q^{i} = (0, q, 0, 0)$ is the heat flux vector which is orthogonal to the velocity vector so that $q^i u_i = 0$. The Einstein's field equations describing the dynamics of the system are then obtained as
\begin{eqnarray}
\rho &=& 3\frac{1}{A^2}\frac{{B_{t}}^2}{B^2} - \frac{1}{B^2}\left( 2\frac{B_{rr}}{B} - \frac{{B_{r}}^2}{B^2} +
\frac{4}{r}\frac{B_{r}}{B} \right), \label{Eq1} \nonumber \\
p &=& \frac{1}{A^2} \left(-2\frac{B_{tt}}{B} - \frac{{B_{t}}^2}{B^2} + 2\frac{A_{t}}{A}\frac{B_{t}}{B} \right) \nonumber \\
&& + \frac{1}{B^2} \left(\frac{{B_{r}}^2}{B^2} + 2\frac{A_{r}}{A}\frac{B_{r}}{B} + \frac{2}{r}\frac{A_{r}}{A} +
\frac{2}{r}\frac{B_{r}}{B} \right),  \label{Eq2}  \\
p &=& -2\frac{1}{A^2}\frac{B_{tt}}{B} + 2\frac{A_{t}}{A^3}\frac{B_{t}}{B} - \frac{1}{A^2}\frac{{B_{t}}^2}{B^2} +
\frac{1}{r}\frac{A_{r}}{A}\frac{1}{B^2} \nonumber \\
&& +  \frac{1}{r}\frac{B_{r}}{B^3} + \frac{A_{rr}}{A}\frac{1}{B^2} - \frac{{B_{r}}^2}{B^4} + \frac{B_{rr}}{B^3},\label{Eq3}\\
q &=& -\frac{2}{AB} \left(-\frac{B_{rt}}{B} + \frac{B_{r}B_{t}}{B^2} + \frac{A_{r}}{A}\frac{B_{t}}{B} \right).\label{Eq4}
\end{eqnarray}
Combining Eqs.~(\ref{Eq2}) and (\ref{Eq3}), we get
\begin{equation}
\displaystyle\frac{A_{rr}}{A} + \displaystyle\frac{B_{rr}}{B} - \left(2 \displaystyle\frac{B_r}{B} +
\displaystyle\frac{1}{r} \right) \left( \displaystyle\frac{A_r}{A} + \displaystyle\frac{B_r}{B} \right) = 0,\label{pi}
\end{equation}
which is the pressure isotropy equation.

The exterior space-time, in the presence an outgoing flux of radiation around the spherically symmetric collapsing matter source, is described by the Vaidya\cite{Vaidya} metric
\begin{equation}
ds_{+}^2 = -\left(1-\frac{2m(v)}{\sf r}\right)dv^2 - 2dvd{\sf r} + {\sf r}^2[d \theta^2 + \sin^2 \theta d
\phi^2], \label{Vm}
\end{equation}
where, the mass function $m(v)$ is a function of the retarded time $v$. Assuming that $\Sigma$ divides the space-times into two distinct regions, the junction conditions joining smoothly the interior space-time (\ref{intm1}) and the exterior space-time (\ref{Vm}) across $\Sigma$ forming the boundary of the star are obtained as\cite{Santos}
\begin{eqnarray}
p_{\Sigma} &=& (q)_{\Sigma}, \label{mp} \\
m_{\Sigma} &=& \Bigg[\frac{r^3 B {B_t}^2}{2 A^2} - r^2 B_r - \frac{r^3 B_r^2}{2 B}\Bigg]_{\Sigma},\label{mm}
\end{eqnarray}
where $m_{\Sigma}$ is the total mass within a sphere of radius $r_{\Sigma}$.

\section{\label{sec3} Generating dynamical solutions}
Note that in Eq.~(\ref{intm1}), the metric potentials $A(r,t)$ and $B(r,t)$ are yet to be specified. To generate a viable dynamical model, let us assume that the system begins its collapse from an initial static configuration ($A_0(r), B_0(r)$). For the initial static configuration, we choose the Pant and Sah\cite{Pant} solution which describes the interior space-time of a static spherically symmetric star in isotropic coordinates. In our construction, we generalize the Pant and Sah\cite{Pant} solution so as to develop a viable model of a collapsing star. We note that Eq.~({\ref{pi}) admits a solution
\begin{eqnarray}
A(r,t) &=& \frac{a(1-\alpha(r) k(t))}{(1+\alpha(r) k(t))}, \label{m1} \\
B(r,t) &=& \frac{(1+\alpha(r) k(t))^2}{(1+r^2/R^2)}, \label{m2}
\end{eqnarray}
for an arbitrary $k(t)$, where
\begin{equation}
\alpha(r) =\sqrt{\frac{1+r^2/R^2}{1+\lambda r^2/R^2}}.\label{alph}
\end{equation}
Obviously, the static limit of the model is obtained by setting $k(t) = K$, a constant (i.e., $\dot{k} =0$). For an evolving system, we need to determine $k(t)$ which can be obtained by solving the junction condition (\ref{mp}). The resultant `surface equation' in this construction turns out to be highly non-linear in nature and extremely difficult to solve. However, it is possible to generate an approximate solution of the equation by setting $k(t) = K + \epsilon h(t)$, with $0 < \epsilon << 1$. Neglecting terms $\cal{O}$$(\epsilon^2)$ and noting that pressure at the boundary of the initial static star vanishes, the surface equation then assumes a simple form
\begin{equation}
\mu \ddot{h} +\nu \dot{h}+\eta h = 0,\label{seq}
\end{equation}
where $\mu$, $\nu$ and $\eta$ are constants evaluated at the boundary $\Sigma$ and are given by
\begin{eqnarray}
\mu &=& -\left[\frac{\alpha (1+K \alpha)^5}{a^2(K \alpha -1)^2}\right]_{\Sigma},\nonumber\\
\nu &=& \left[\frac{\alpha(1+K \alpha)^2\sqrt{(1-\alpha^2)(\alpha^2\lambda-1)}}{a R(K\alpha-1)^2}\right]_{\Sigma},\nonumber\\
\eta &=& \left[\frac{2\alpha(2-3K \alpha+\lambda K\alpha^5(1+2K \alpha(K\alpha-1)))}{R^2(k^2\alpha^2-1)^2}\right]_{\Sigma}.
\end{eqnarray}
Eq.~(\ref{seq}) is easily solvable and the most general solution of the equation can be written as
\begin{equation}
h(t) = C e^{\frac{t}{2\mu}\left(-\nu - \sqrt{\nu^2-4\mu\eta}\right)} + D e^{\frac{t}{2\mu}\left(-\nu + \sqrt{\nu^2-4\mu\eta}\right)},\label{sol}
\end{equation}
where $C$ and $D$ are integration constants.

We assume that the collapse begins in the remote past $t\rightarrow -\infty$ from an initial static configuration as the star loses its equilibrium. This implies that we must have $k(t\rightarrow -\infty) = K$, where $K$ is a constant as described in the static Pant and Sah\cite{Pant} model. For a collapsing (contracting) sphere, without any loss of generality, we set $D=0$ which ensures that $\dot{k},~\dot{h} < 0$. On top of this, for real values of $h(t)$, we need to fix the model parameters so that the condition $\nu^2 \geq 4\mu\eta$ is satisfied.

Subsequently, from Eqs.~(\ref{Eq1})-(\ref{Eq4}), the energy-density, pressure and heat flux density are obtained as
\begin{eqnarray}
\rho(r,t) &=& \rho_s(r) + \frac{12\epsilon\alpha\left[-5+\lambda\alpha^4(1-4K \alpha)\right]h(t)}{R^2(1+K\alpha)^6},\label{mEq1} \\
p(r,t) &=& p_s(r) + \frac{\epsilon\alpha\left[8a^2\left[2-3K\alpha+\lambda K\alpha^5\left[1+2K \alpha(K\alpha-1)\right]\right]h+R^2(1+K\alpha)^7\ddot{h}\right]}{a^2R^2(K\alpha-1)^2(K\alpha+1)^6},\label{mEq2}\\
q &=& \frac{4\alpha\epsilon\sqrt{(\alpha^2-1)(1-\lambda\alpha^2)}\dot{h}}{a R\left[\alpha^2(\epsilon h +K)^2-1\right]^2}.\label{mEq3}
\end{eqnarray}
In Eqs.~(\ref{mEq1}) and (\ref{mEq2})
\begin{eqnarray}
\rho_s(r) &=& \frac{12(1 + \lambda K\alpha^5)}{R^2(1+K\alpha)^5},\label{smEq1} \\
p_s(r) &=& \frac{4(\lambda K^2\alpha^6-1)}{R^2(1-K\alpha)(1+K\alpha)^5},\label{smEq2}
\end{eqnarray}
correspond to the density and pressure of the initial static configuration.

From (\ref{mm}), the total mass $m(r,t)$ within a radius $r \leq r_{\Sigma}$ at any instant $t$ is obtained as
\begin{equation}
m(r,t) = m_s(r)+ \frac{2\epsilon R (1+\lambda+2K\alpha^3\lambda)\left[(\alpha^2-1)(1-\alpha^2\lambda)\right]^{\frac{3}{2}}h(t)}{\alpha^3(1-\lambda)^3},\label{masseq2}
\end{equation}
where,
\begin{equation}
m_s (r) = \frac{2R(1+K\alpha^3)(1+K\alpha^3\lambda)\left[(\alpha^2-1)(1-\alpha^2\lambda)\right]^{\frac{3}{2}}}{\alpha^6(1-\lambda)^3},\label{smass}
\end{equation}
is the mass of the initial static configuration.

Let $b$ be the radial distance where the pressure of the initial static star vanishes, i.e., $p_s(r=b) = 0$. The physical radius $b_0$ of the initial static star can then be obtained from the relation (see \cite{Deb})
\begin{equation}
b_0 = b\left(1+\frac{m_s(b_0)}{2b}\right)^2.\label{radeq}
\end{equation}
Matching the static interior solution to the Schwarzschild exterior at the boundary, we obtain
\begin{eqnarray}
\frac{(1+K \alpha_b)^2}{(1+b^2/R^2)} &=& \left(1+\frac{m_s(b)}{2b}\right)^2,\label{sjc1}\\
K &=& \frac{1}{\sqrt{\lambda}\alpha_b^3},\label{sjc2}
\end{eqnarray}

\begin{eqnarray}
[(1+a)^4+K^4(1-a)^4-8(1+a)^2+16-2K^2(1-a^2)^2-8K^2(1-a)^2] \nonumber\\
+ [2\lambda(1+a)^4-16\lambda(1+a)^2-8(1+a)^2+32(1+\lambda) -2K^2(1+\lambda)(1-a^2)^2\nonumber\\
-8K^2(2+\lambda)(1-a)^2+2K^4(1-a)^4]y^2+[\lambda^2(1+a)^4-8\lambda^2(1+a)^2\nonumber\\
-8\lambda(1+a)^2(1+4\lambda+\lambda^2)-2\lambda K^2(1-a^2)^2-8K^2(1-a)^2(1+2\lambda)\nonumber\\
+K^4(1-a)^4]y^4-[8\lambda^2(1+a)^2-32(1+\lambda)-8\lambda K^2(1-a)^2]y^6+16\lambda^2 y^8 = 0,\label{sjc3}
\end{eqnarray}
where,
$$ y = \frac{b}{R},~~~~~~\alpha_b = \sqrt{\frac{1+y^2}{1+\lambda y^2}}.$$
Eqs.~(\ref{radeq})-(\ref{sjc3}) can be utilized to fix the model parameters for a specific choice of mass and radius of the initial static star.

\section{\label{sec4} Physical analysis}

In order to analyze the physical behaviour of the collapsing model, let us now assume that the star begins its collapse from an initial static configuration of mass $m_s = 5~M_{\odot}$ and physical radius $b_0 = 15~$km. Setting $\lambda =20$, we determine the model parameters as $b = 4.7808~$km, $R = 10.5828~$km, $a = 4.5958$ and $K = 1.9386$. Assuming $C = -100$ and $\epsilon = 0.1$, variations of $k(t)$ and $\dot{k}$ have been shown in Fig.~(\ref{fg1}) and (\ref{fg2}), respectively. Evolution of total mass $m(r_{\Sigma},t)$, proper radius $(Br)_{\Sigma}$, heat-flux $q$ and surface density $\rho({r_{\Sigma},t})$ of the collapsing matter have been shown in Fig.~(\ref{fg3}) -(\ref{fg6}), respectively. With the evolution of time, the total mass $m(r_{\Sigma},t)$ starts decreasing from its initial value of $5~M_{\odot}$ and eventually becomes zero over a finite time (Fig.~\ref{fg3}). Similarly, the radius also shrinks to zero over a finite time (Fig.~\ref{fg4}). Though, in this model,  the ratio $2m(r,t)/(rB)$ is time dependent; it is interesting to note that singularity is never formed and the stellar mass completely evaporates over a finite time. Similar results may be found in Ref.~\cite{Pinheiro1}.

\section{\label{sec5}Thermal evolution}

Let us now investigate the thermal evolution of the collapsing system generated in Sec.~\ref{sec3}. It is well known that the Eckart formalism of thermodynamics suffers many pathologies, some of which include super-luminal propagation velocities for the dissipative fluxes as well as the prediction of unstable equilibrium states\cite{anile}. Gravitational collapse of a stellar object is dissipative in nature and is usually accompanied by heat generation via neutrino emission or free-streaming radiation. Various investigations have shown that relaxational effects predict higher core temperatures and significantly different luminosity profiles when compared to their non-causal counterparts\cite{gov1,gov2,nolene,megan}.  In order to study the impact of the relaxational effects brought about by the heat flow, we will employ the truncated causal heat transport equation\cite{Thiru1}
\begin{equation}
{\tau} h_i{}^j {\dot q}_{j} + q_i = -\kappa(h_i{}^j \nabla_j T + T {\dot u}_i),\label{cmc}
\end{equation}
where $\kappa$ is the thermal conductivity, ${\dot u}_j =  u_{j;i}u^i$ and $h_{ij} = g_{ij} + u_iu_j$ is the projection tensor and $\tau$ is the relaxation time. We obtain the Eckart temperature by setting $\tau = 0$ in (\ref{cmc}). We assume that the  neutrinos are thermally generated within the stellar core with energies of the order of $k_{B}T$. At neutron star densities, neutrino trapping takes place via electron-neutrino scattering and nucleon absorption. The mean collision time for thermally generated neutrinos is given by
\begin{equation} \label{tauu}
\tau_{\rm c} \propto T^{-3/2},
\end{equation}
to good approximation\cite{marti}. Following (\ref{tauu}) we adopt a power-law dependence for the thermal conductivity and relaxation time:
\begin{equation}
\kappa = \gamma T^3{\tau}_{\rm c}, \hspace{2cm} \tau_{\rm c}
=\left(\frac{\alpha}{\gamma}\right) T^{-\sigma}, \label{a28}
\end{equation}
where $\alpha \geq 0$, $\gamma \geq 0$ and $\sigma \geq 0$ are constants. We further assume that the relaxation time is directly proportional to the mean collision time
\begin{equation}
\tau = \left(\frac{\beta \gamma}{\alpha}\right) \tau_{\rm c}
\label{a30}\,
\end{equation}
where $\beta$ ($\geq 0$) is a constant. The causal transport equation (\ref{cmc}) together with the above assumptions reduces to
\begin{equation}
\beta (qB)^{\dot{}} T^{-\sigma} + A (q B) = - \alpha
\frac{T^{3-\sigma} (AT)'}{B} \label{temp1} \,
\end{equation} where the Eckart temperature $T_0$ is obtained by setting $\beta = 0$ in (\ref{temp1}). In the case of constant collision time $(\sigma = 0)$ we are in a position to write down the solution to (\ref{temp1}) as
\begin{equation}
(AT)^4 = - \frac{4}{\alpha} \left[\beta\int A^3 B
(qB)_{,t}{\mathrm d} r + \int A^4 q B^2 {\mathrm d} r\right] +
F(t), \label{caus0}
\end{equation}
where $F(t)$ is an arbitrary function of integration. The function ${F}(t)$ can be determined from the effective surface temperature of a star as given by
\begin{equation} \label{f10}
\left({T^4}\right)_{\Sigma} = \left(\frac{1}{r^2B^2}\right)_{\Sigma}\left(\frac{L_{\infty}}{4\pi\delta}\right),
\end{equation}
where $L_{\infty}$ is the total luminosity at infinity and $\delta$ ($>0$) is a constant.

Making use of the solution generated in Sec.~\ref{sec3} and for the particular case considered in Sec.~\ref{sec4}, we have plotted the temperature within the collapsing stellar core as a function of time in Fig.~(\ref{fg7})-(\ref{fg10}). Fig.~(\ref{fg7}) exhibits the non-causal temperature at the centre of the collapsing star. As expected, the temperature decreases monotonically with time. Fig.~(\ref{fg8}) shows the causal temperature profile at the centre of the star as a function of time. A comparison of Fig.~(\ref{fg7}) and (\ref{fg8}) shows that the causal temperature is always greater than the non-causal temperature. A comparison of Fig.~(\ref{fg9}) and (\ref{fg10}) clearly demonstrates that the casual temperature dominates its non-causal counterpart at the surface as the collapse proceeds. It is clear that relaxational effects contribute to the enhancement of the temperature at each interior point of the collapsing star. It is interesting to note that even though our model is based on a weak heat flux approximation $(0 < \epsilon << 1)$, relaxational effects lead to very different outcomes for the temperature.

\section{\label{sec6}Discussions}

In our work, we have generated a dynamical solution from the static stellar model of Pant and Sah\cite{Pant} to investigate the nature of dissipative collapse. This has been achieved by allowing a constant parameter in the static model to evolve with time. The resulting dynamical model is a radiating collapsing star with heat conduction enveloped by a radiation atmosphere. Though in our construction, the star begins its collapse from an initial static configuration described by the Pant and Sah\cite{Pant} model, unlike many previous models describing collapse from an initial static configuration, the usual method of assuming metric separability in their variables $r$ and $t$ has not been adopted in our approach. Secondly, the back ground space-time has been couched in spatially isotropic coordinates. The static model of Pant and Sah\cite{Pant}, as analyzed by Deb {\em et al}\cite{Deb}, has the following key features: (1) $\rho >0$, $p > 0$; (2) $\rho' < 0$, $p' < 0$; (3) $dp/d\rho < 1$. This implies that the collapse begins from a physically acceptable initial configuration which includes the fulfillment of (at least) the weak energy condition. The collapse is found to proceed without formation of an event horizon as it radiates all its mass energy over a finite time. We have also studied the thermodynamics of the collapsing star within the framework of extended irreversible thermodynamics. Our results confirm earlier findings (through various different approaches) that relaxational effects can significantly alter the physical characteristics such as the temperature of the collapsing system.

\begin{acknowledgments}
RS gratefully acknowledge support from the Inter University Centre for Astronomy and Astrophysics (IUCAA), Pune, India, under its Visiting Research Associateship Programme.
\end{acknowledgments}


\begin{thebibliography}{99}

\bibitem{Shapiro} S. L. Shapiro and S. A.  Teukolosky, {\em Black Holes, White Dwarfs and Neutron Stars: The Physics of Compact Objects}, Wiley, New York  (1983).
\bibitem{Finch} M. R. Finch and J. E. F. Skea, {\em http://www.dft.if.uerj.br/users/JimSkea/papers/pfrev.ps} (1998).
\bibitem{Delgaty} M. S. R. Delgaty and K. Lake, {\it Computer Physics Communications} {\bf 115}, 395 (1998).
\bibitem{Tikekar} R. Tikekar, {\it J. Math. Phys.} {\bf 31}, 2454 (1990).
\bibitem{Sharma01} R. Sharma, S.  Mukherjee and S. D. Maharaj, {\em Gen. Relativ. Grav.} {\bf33}, 999 (2001).
\bibitem{Karma} S. Karmakar, S. Mukherjee, R. Sharma and S. D. Maharaj, {\em Pramana - J. Phys.} {\bf68}, 881 (2007).
\bibitem{Sharma02} R. Sharma, S. Mukherjee, M. Dey and J. Dey, {\it Mod. Phys. Lett. A} {\bf 17}, 827 (2002).
\bibitem{Sharma03} R. Sharma and S.  Mukherjee, {\em Mod. Phys. Lett. A} {\bf16}, 1049 (2001).
\bibitem{singe1} G. P. Singh and S. Kotambkar, {\em Pramana - J. Phys.} {\bf 65}, 35 (2005).
\bibitem{singe2} L. K. Patel and G. P. Singh, {\em Gravitation and Cosmology} {\bf 7}, 52 (2001).
\bibitem{Pant} D. N. Pant and A. Sah, {\it Phys. Rev. D} {\bf32}, 1358 (1985).
\bibitem{Buchdahl} H. A. Buchdahl, {\it Astrophys. J.} {\bf 140}, 1512 (1964).
\bibitem{Deb} R. Deb, B. C. Paul and R. Tikekar, {\it Pramana - J. Phys.} {\bf79}, 211 (2012).
\bibitem{OppenS} J. R. Oppenheimer and H. Snyder, {\it Phys. Rev.} {\bf56}, 455 (1939).
\bibitem{Vaidya} P. C. Vaidya, {\it Nature} {\bf 171}, 260 (1953).
\bibitem{Santos} N. O. Santos, {\it Mon. Not. R. Astron. Soc.} {\bf 216}, 403 (1985).
\bibitem{Herrera1} L. Herrera, J. Ospino and A. Di Prisco, {\it Phys. Rev. D} {\bf77}, 027502 (2008).
\bibitem{Herrera2} L. Herrera, N. O. Santos and A. Wang, {\it Phys. Rev. D} {\bf 78}, 084026 (2008).
\bibitem{Herrera3} L. Herrera, A. di Prisco, J. L. Hern$\acute{a}$ndez-Pastora and N. O. Santos, {\it Phys. Lett. A} {\bf 237}, 113 (1998).
\bibitem{Herrera4} L. Herrera and N. O. Santos, {\it Mon. Not. R. Astron. Soc.} {\bf287}, 161 (1997).
\bibitem{Herrera5} L. Herrera, A. di Prisco, J. Martin, J. Ospino, N. O. Santos and O. Troconis, {\it Phys. Rev. D} {\bf69}, 084026 (2004).
\bibitem{Herrera6} L. Herrera and W. Barreto, {\it Int. J. Mod. Phys. D} {\bf 20}, 1265 (2011).
\bibitem{Herrera7} L. Herrera and N. O. Santos, {\it Phys. Rep.} {\bf286}, 53 (1997).
\bibitem{Herrera8} L. Herrera, J. Martin and J. Ospino, {\it J. Math. Phys.} {\bf43}, 4889 (2002).
\bibitem{Herrera9} L. Herrera, {\it Int. J. Mod. Phys. D} {\bf20}, 1689 (2011).
\bibitem{Bonnor} W. B. Bonnor, A. K. G. de Oliveira and N. O. Santos {\it Phys. Rep.} {\bf 181}, 269 (1989).
\bibitem{Oliveira1} A. K. G. de Oliveira and N. O. Santos, {\it Astrophys. J.} {\bf 312}, 640 (1987); {\it Mon. Not. R. Astron. Soc.} {\bf216}, 1001 (1988).
\bibitem{Oliveira2} A. K. G. de Oliveira, C. A. Kolassis and N. O. Santos, {\it Mon. Not. R. Astron. Soc.} {\bf231}, 1011 (1988).
\bibitem{Prisco1} A. Di Prisco, L. Herrera, G. L. Denmat, M. A. H. MacCallum and N. O. Santos, {\it Phys. Rev. D} {\bf 76}, 064017 (2007).
\bibitem{Prisco2} A. Di Prisco, L. Herrera and V. Varela, {\it Gen. Relativ. Grav.} {\bf29}, 1239 (1997).
\bibitem{Pinheiro1} G. Pinheiro and R. Chan, {\it Gen. Relativ. Grav.} {\bf 40}, 2149 (2008); {\it ibid} {\bf 43}, 1451 (2011); {\it ibid} {\bf 45}, 243 (2013).
\bibitem{Chan1} R. Chan, {\it Mon. Not. R. Astro.} {\bf 316}, 588 (2000); {\it Astron. \& Astrophys.} {\bf 368}, 325 (2001).
\bibitem{Chan2} R. Chan, M. F. A. da Silva and J. F. V. da Rocha, {\it J. Math. Phys. D} {\bf 12}, 347 (2003).
\bibitem{Chan3} R. Chan, L. Herrera and N. O. Santos {\it Mon. Not. R. Astron. Soc.} {\bf267}, 637 (1994).
\bibitem{Tikekar1} R. Tikekar and L. K. Patel, {\it Pramana - J. Phys.} {\bf 39}, 17 (1992).
\bibitem{Maharaj1} S. D. Maharaj and M. Govender, {\it Pramana - J. Phys.} {\bf 54}, 715 (2000).
\bibitem{Sharif1} M. Sharif and K. Iqbal, {\it Mod. Phys. Lett. A} {\bf24}, 1533 (2009).
\bibitem{Sharif2} M. Sharif and A. Siddiqe, {\it Mod. Phys. Lett. A} {\bf25}, 2831 (2010); {\it Gen. Relativ. Grav.} {\bf43},  73 (2011).
\bibitem{Sharif3} M. Sharif and G. Abbas, {\it J. Phys. Soc. Jpn.} {\bf80}, 104002 (2011); {\it Astrophys. Space Sci.} {\bf335}, 515 (2011).
\bibitem{Sharif4} M. Sharif and F. Sundas, {\it Gen. Relativ. Grav.} {\bf43}, 127 (2011).
\bibitem{Sharif5} M. Sharif and S. Fatima, {\it Gen. Relativ. Grav.} {\bf43}, 127 (2011).
\bibitem{Barreto1} W. Barreto, B. Rodr$\acute{i}$guez, L. Rosales and O. Serrano, {\it Gen. Relativ. Grav.} {\bf 39}, 23 (2007).
\bibitem{Ghezzi1} C. Ghezzi, {\it Phys. Rev. D} {\bf72}, 104017 (2005).
\bibitem{Goswami1} R. Goswami and P. S. Joshi, {\it Class. Quantum Grav.} {\bf21}, 3645 (2004); {\it Class. Quantum Grav.} {\bf19}, 0129 (2002); {\it Phys. Rev. D} {\bf69}, 027502 (2004).
\bibitem{Thiru1} S. Thirukkanesh and S. D. Maharaj, {\it Math. Meth. Appl. Sci.} {\bf 32}, 684 (2009).
\bibitem{Govinder1} K. S. Govinder, M. Govender and R. Maartens, {\it Mon. Not. R. Astron. Soc.} {\bf 299}, 809 (1998).
\bibitem{Govender1} M. Govender and S. Thirukkanesh, {\it Int. J. Mod. Phys.} {\bf 48}, 3558 (2009).
\bibitem{Goenner1} D. Sch$\ddot{a}$fer and H. F. Goenner, {\it Gen. Relativ. Grav.} {\bf 42}, 2119  (2000).
\bibitem{Sarwe1} S. Sarwe and R. Tikekar, {\em Int. J. Mod. Phys. D} {\bf19}, 1889 (2010).
\bibitem{Sharma1} R. Sharma and R. Tikekar, {\it Gen. Relativ. Grav.} {\bf44}, 2503 (2012); {\it Pramana- J. Phys.} {\bf79}, 501 (2012).
\bibitem{t1g1} S. Thirukannesh and M. Govender, {\em Int. J. Mod. Phys. D} {\bf 22}, 1350087 (2013).
\bibitem{mks} M. Govender, K. P. Reddy and S. D. Maharaj, {\em Int. J. Mod. Phys. D} to appear (2013).
\bibitem{Joshi01} P. S. Joshi and Malafarina, {\it Int. J. Mod. Phys. D} {\bf20}, 2641 (2011).
\bibitem{Kramer1} D. Kramer, {\it J. Math. Phys.} {\bf 33},
1458 (1992).
\bibitem{Maharaj2} S. D. Maharaj and M. Govender, {\it Aust. J. Phys.} {\bf 50}, 959 (1997).
\bibitem{Chandra} S. Chandrasekhar, {\it Observatory} {\bf57}, 373 (1934).
\bibitem{anile} A. M. Anile, D. Pavon and V. Romano, ``The case for hyperbolic theories of dissipation in relativistic fluids." Report gr-qc/9810014.
\bibitem{gov1} M. Govender, S. D. Maharaj and R. Maartens, {\it
Class. Quantum Grav.} {\bf 15},323(1998).
\bibitem{gov2}M. Govender, R. Maartens and S. D. Maharaj, {\it
Mon. Not. R. Astron. Soc.} {\bf 310},557 (1999).
\bibitem{nolene} N. F. Naidu, M. Govender and K. S. Govinder, {\it Int. J. Mod. Phys.} D {\bf 15}, 1053 (2006).
\bibitem{megan} M. Govender, {\it Int. J. Mod. Phys. D} {\bf {22}} , 1350049 (2013).
\bibitem{marti} J. Martinez, {\it Phys. Rev. D}  {\bf {53}}, 6921 (1996).
\bibitem{Thiru2} S. Thirukannesh, S. S. Rajah and S. D. Maharaj, {\it J. Math. Phys.} {\bf {53}}, 032506 (2012).

\end{thebibliography}


\newpage

\begin{figure}
\includegraphics[width=0.75\textwidth]{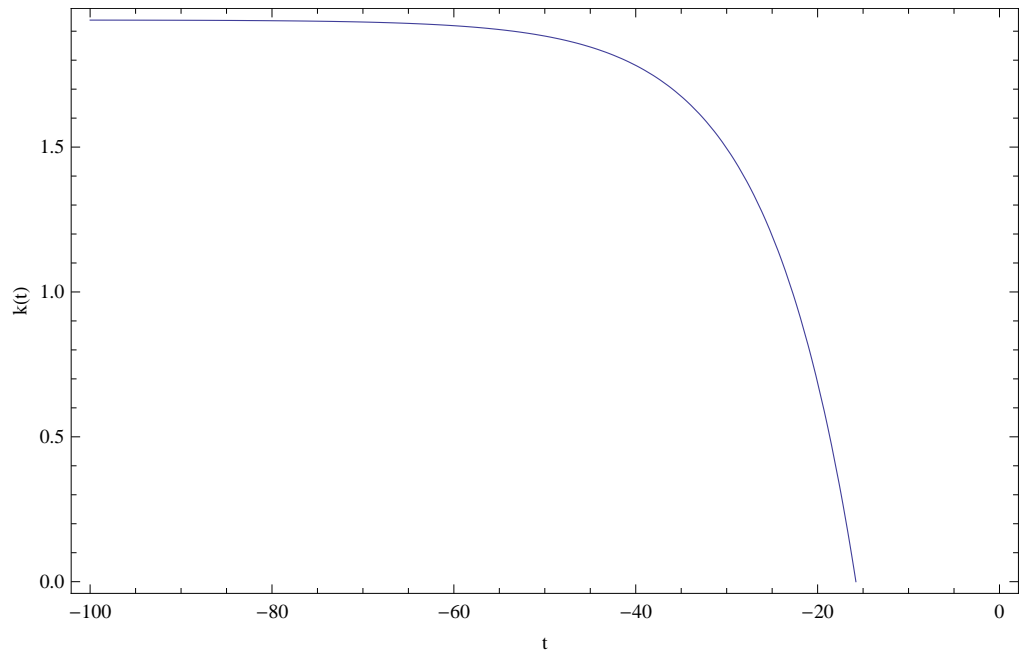}\caption{Evolution of $k(t)$.} \label{fg1}
\end{figure}

\begin{figure}
\includegraphics[width=0.75\textwidth]{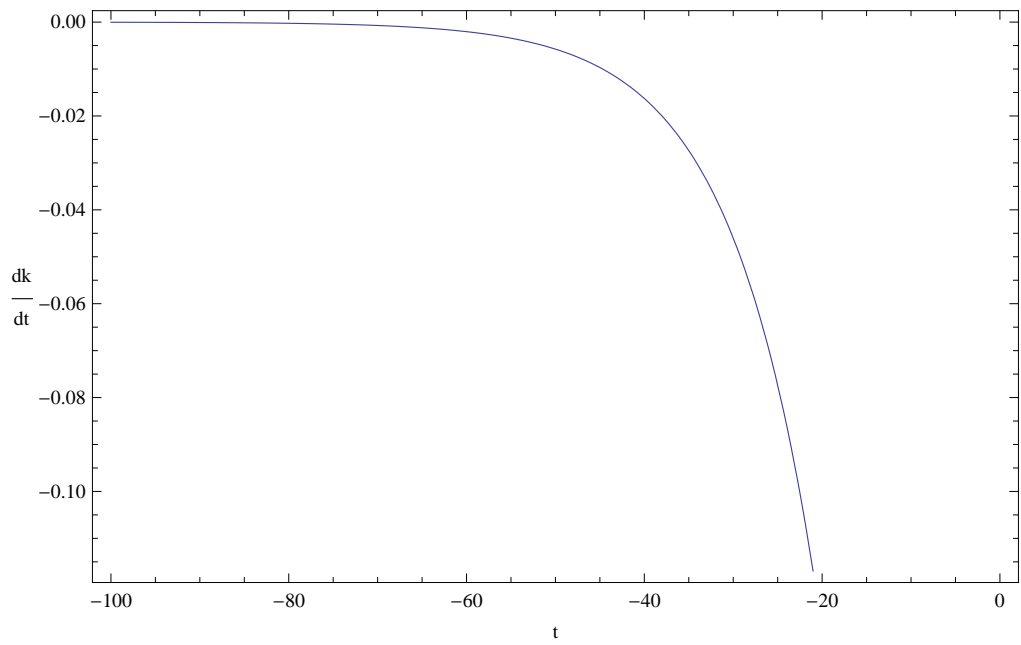}\caption{Evolution of $\dot{k}$.} \label{fg2}
\end{figure}

\begin{figure}
\includegraphics[width=0.75\textwidth]{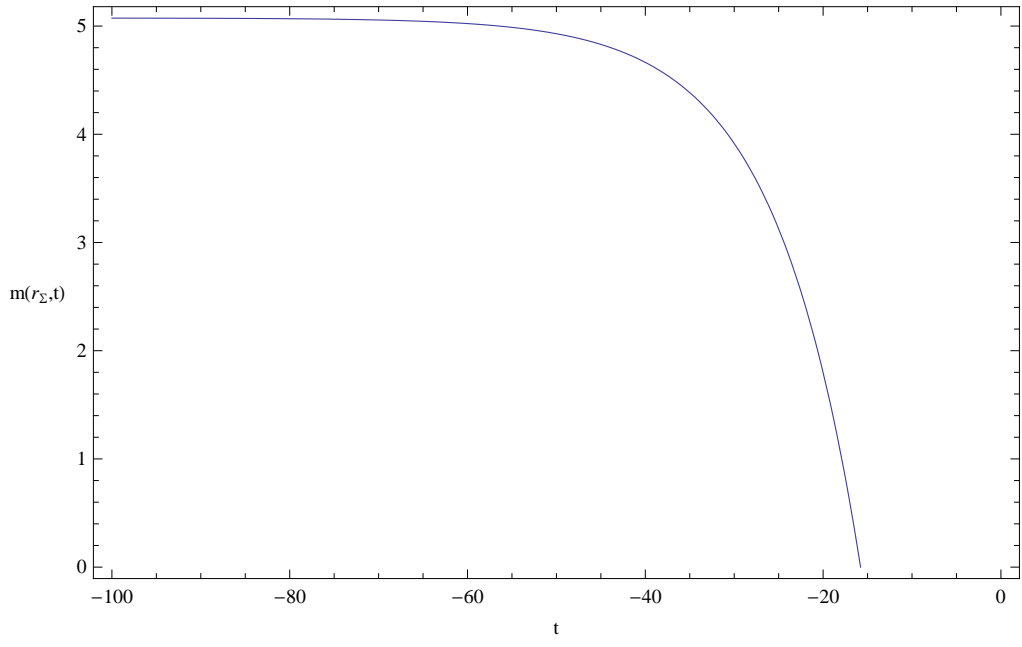}\caption{Evolution of mass $m(r_{\Sigma},t)$.} \label{fg3}
\end{figure}

\begin{figure}
\includegraphics[width=0.75\textwidth]{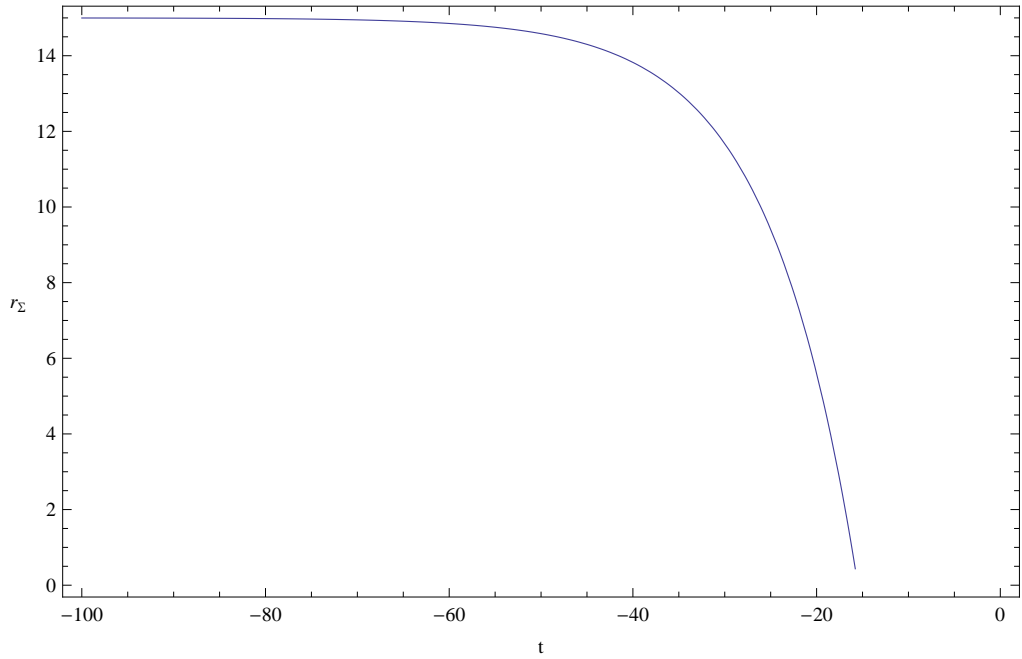}\caption{Evolution of physical radius $r_{\Sigma}B(r,t)$.} \label{fg4}
\end{figure}

\begin{figure}
\includegraphics[width=0.75\textwidth]{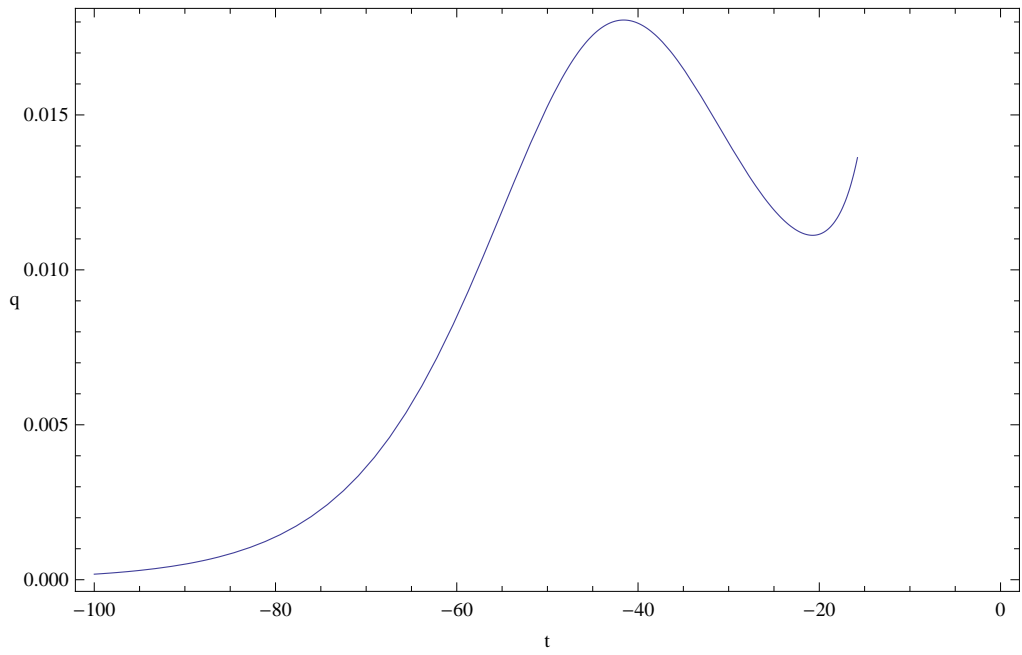}\caption{Evolution of heat flux $q(r_{\Sigma},t)$.} \label{fg5}
\end{figure}

\begin{figure}
\includegraphics[width=0.75\textwidth]{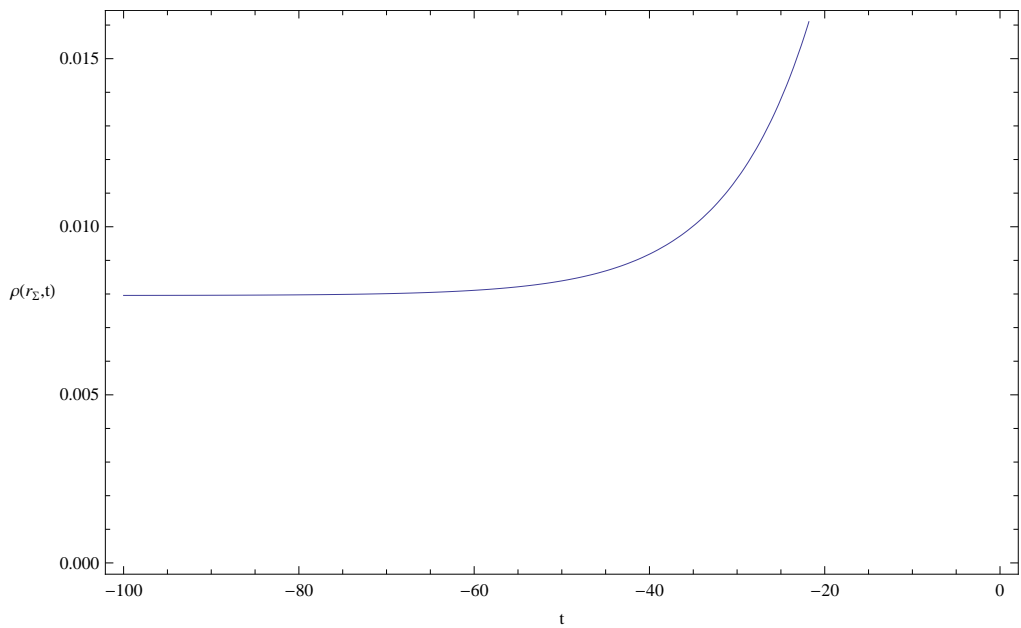}\caption{Evolution of surface density $\rho(r_{\Sigma},t)$.} \label{fg6}
\end{figure}

\begin{figure}
\includegraphics[width=0.75\textwidth]{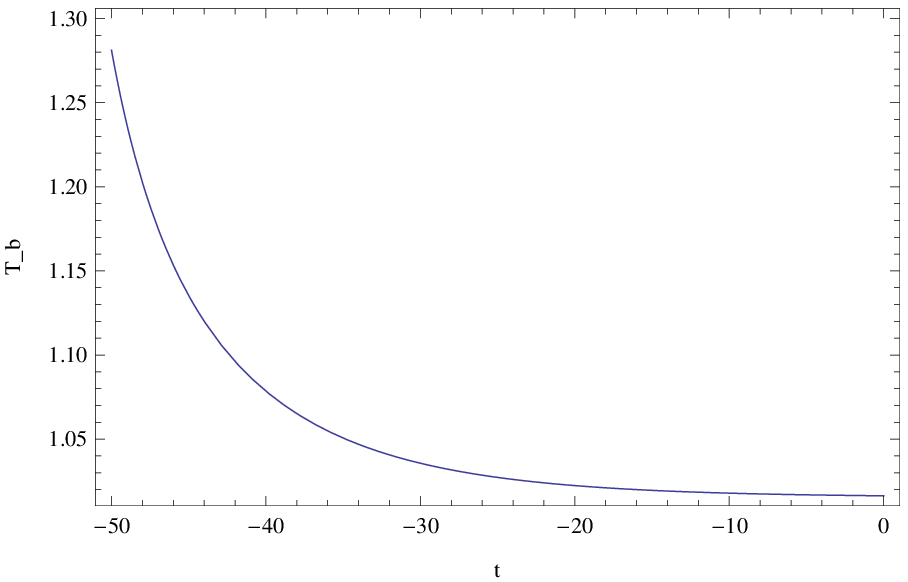}\caption{Non-causal temperature profile at the centre} \label{fg7}
\end{figure}

\begin{figure}
\includegraphics[width=0.75\textwidth]{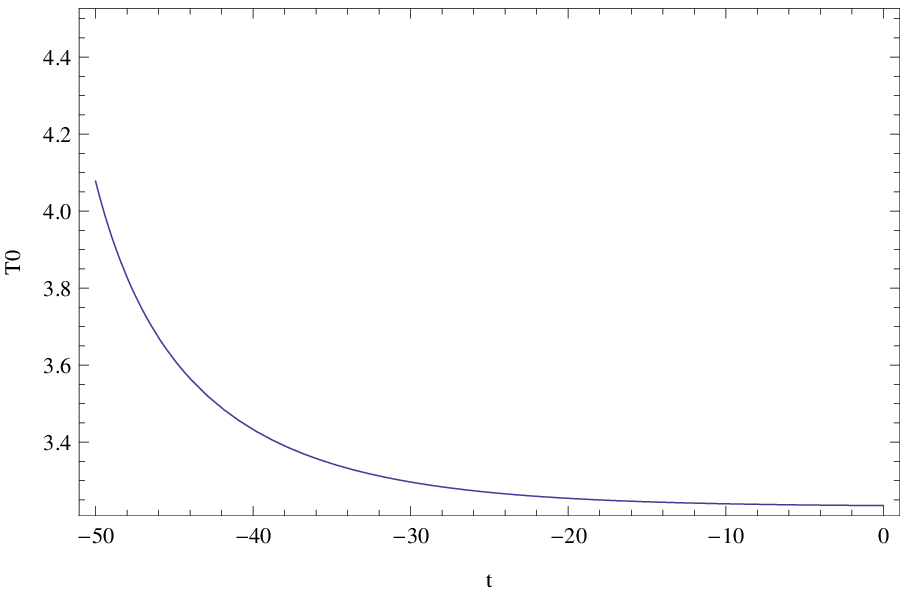}\caption{Causal temperature profile at the centre} \label{fg8}
\end{figure}

\begin{figure}
\includegraphics[width=0.75\textwidth]{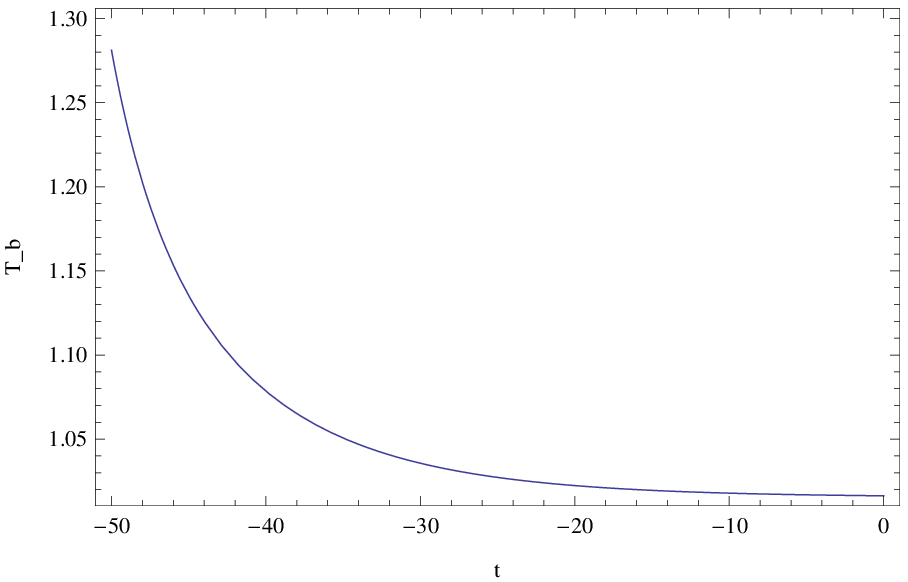}\caption{Non-causal temperature profile at the surface} \label{fg9}
\end{figure}

\begin{figure}
\includegraphics[width=0.75\textwidth]{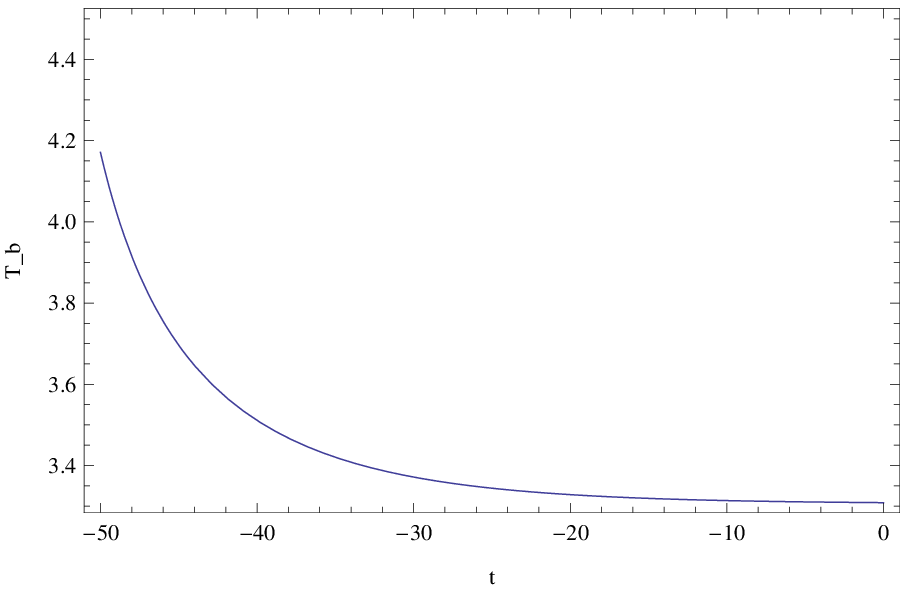}\caption{Causal temperature profile at the surface} \label{fg10}
\end{figure}

\end{document}